\documentclass[10pt, conference]{IEEEtran}
\IEEEoverridecommandlockouts
\usepackage{xspace}
\usepackage{textgreek}
\usepackage{balance}
\usepackage{graphicx}
\usepackage[hyphens]{url}
\usepackage{epstopdf}
\usepackage{microtype}
\usepackage{booktabs}
\usepackage[table,xcdraw]{xcolor}
\usepackage{caption}
\usepackage{subcaption}
\usepackage{amsmath}
\usepackage{enumitem}
\usepackage{amssymb}
\usepackage{pifont}
\usepackage{comment}
\newcommand{\RN}[1]{%
  \textup{\uppercase\expandafter{\romannumeral#1}}%
}
\iftrue 
	
\else
	
\fi

\newcommand{\PreserveBackslash}[1]{\let\temp=\\#1\let\\=\temp}
\newcolumntype{M}[1]{>{\PreserveBackslash\centering}m{#1}}
\newcolumntype{C}[1]{>{\PreserveBackslash\centering}p{#1}}
\newcolumntype{R}[1]{>{\PreserveBackslash\raggedleft}p{#1}}
\newcolumntype{L}[1]{>{\PreserveBackslash\raggedright}p{#1}}

\newcommand{\fakepar}[1]{\smallbreak\noindent{}}
\newcommand{\boldpar}[1]{\smallbreak\noindent\textbf{#1.}}

\newcommand{\ieee}{\mbox{IEEE~802.15.4}\xspace}

\newcommand{\usdn}{\mbox{$\mu$SDN}\xspace}
\newcommand{\wise}{\mbox{SDN-WISE}\xspace}

\usepackage{manyfoot}
\DeclareNewFootnote{A}
\DeclareNewFootnote{B}

\iftrue
    \newcommand{\mike}[1]{\footnoteB{{\color{red}\bf MB: #1}\color{red}}}
    \newcommand{\israat}[1]{\footnoteB{{\color{red}\bf IH: #1}\color{red}}}
    \newcommand{\miheer}[1]{\footnoteA{{\color{blue}\bf MK: #1}\color{blue}}}
\else
    \newcommand{\mike}[1]{}
    \newcommand{\israat}[1]{}
    \newcommand{\miheer}[1]{}
\fi

\makeatletter
\def\ps@IEEEtitlepagestyle{%
	\def\@oddfoot{\mycopyrightnotice}%
	\def\@evenfoot{}%
}
\def\mycopyrightnotice{%
	\begin{minipage}{\textwidth}
		\centering\tiny{978-1-6654-0522-5/21/\$31.00~\copyright{} 2021 IEEE.  Personal use of this material is permitted.  Permission from IEEE must be obtained for all other uses, in any current or future media, including reprinting/republishing this material for advertising or promotional purposes, creating new collective works, for resale or redistribution to servers or lists, or reuse of any copyrighted component of this work in other works.}
	\end{minipage}
	\gdef\mycopyrightnotice{}
}

\begin{document}

\title{
Embedded vs. External Controllers in Software-Defined IoT Networks
}

\iffalse
\author{
\IEEEauthorblockN{Anonymous Author(s)} 
}
\else
\author{
\IEEEauthorblockN{Miheer Kulkarni\IEEEauthorrefmark{1}, 
Michael Baddeley\IEEEauthorrefmark{2}, 
and
Israat Haque\IEEEauthorrefmark{1}} \vspace{-2.50mm}\\ 
\IEEEauthorblockA{\IEEEauthorrefmark{1}Department of Computer Science, Dalhousie University, Halifax, Canada -- \url{{miheer; israat}@dal.ca} }
\IEEEauthorblockA{\IEEEauthorrefmark{2}Secure Systems Research Center (SSRC), Technology Innovation Institute (TII), Abu Dhabi, UAE -- \url{michael@ssrc.tii.ae} } \vspace{-5.00mm}\\
}
\fi

\maketitle


\begin{abstract}
The flexible and programmable architectural model offered by Software-Defined Networking (SDN) has re-imagined modern networks. Supported by powerful hardware and high-speed communications between devices and the controller, SDN provides a means to virtualize control functionality and enable rapid network reconfiguration in response to dynamic application requirements. However, recent efforts to apply SDN's centralized control model to the Internet of Things (IoT) have identified significant challenges due to the constraints faced by embedded low-power devices and networks that reside at the IoT edge. In particular, reliance on external SDN controllers on the backbone network introduces a performance bottleneck (e.g., latency). To this end, we advocate a case for supporting Software-Defined IoT networks through the introduction of lightweight SDN controllers \emph{directly} on the embedded hardware. We firstly explore the performance of two popular SDN implementations for IoT mesh networks, $\mu$SDN and SDN-WISE, showing the former demonstrates considerable gains over the latter. We consequently employ $\mu$SDN to conduct a study of embedded vs. external SDN controller performance. We highlight how the advantage of an embedded controller is reduced as the network scales, and quantify a point at which an external controller should be used for larger networks. 
\end{abstract}

\section{Introduction}
\label{sec:introduction}

There has been a significant effort from the networking community to harness the flexibility of Software-Defined Networking (SDN)~\cite{ietf_sdn}, which lifts the control plane from individual devices, centralizes decision making, and encourages dynamic network configuration through the provision of open APIs on commodity hardware. SDN has typically been applied to the wired and optical networks found in cloud computing use-cases, but has also found a home in wireless communications~\cite{haque2016wireless}. More recent efforts have tried to extend the SDN concept to the Internet of Things (IoT) networks, where SDN`s programmable networking model could help support research challenges such as adapting to dynamic application requirements and providing end-to-end Quality-of-Service (QoS) guarantees for IoT communications~\cite{baddeley2020thesis}. However, Software-Defined IoT networks face a significant hurdle. Although supported by cloud and edge services, IoT networks typically consist of low-power embedded devices which face lossy wireless channels and limited device and computing resources. Furthermore, industrial applications often consist of large multi-hop wireless sensor and actuator networks with hundreds of nodes. In such scenarios, SDN`s fully-centralized networking model is no longer viable: concessions and trade-offs must be made~\cite{baddeley2018evolving}. Crucially, reliance on an externally situated controller on a backbone network introduces an additional and constant overhead to SDN control messaging.

To address this challenge, we advocate embedding an SDN controller directly within a low-power wireless network and virtualizing basic control functions on constrained devices. By removing reliance on an external controller implementation, this significantly reduces latencies between the controller and embedded SDN devices. Furthermore, modern low-power wireless chipsets are considerably more powerful than prior generations, and are more than capable of hosting simple control functions. Yet while there are a number of publicly available SDN architectures for IoT~\cite{gallucio2015sdn,theodorou2017coralsdn,alves2017itsdn}, only the Coniki-NG based \usdn~\cite{baddeley2018evolving} has (to-date) implemented an embedded controller. Furthermore, although \usdn specifies it has taken this approach, there exists no subsequent study specifically evaluating its performance against an external controller. 

\boldpar{Our contributions} We show how an embedded SDN controller can achieve significant gains in lower bit-rate networks, and quantify the performance of an embedded vs. external controller as the network scales. We observe a `crossover' point and highlight the scenarios where one controller outperforms the other. Specifically, as the network scales, traffic load has a significant effect on the performance of an embedded controller -- negating earlier latency gains. Thus, application developers can choose embedded vs. external controller based on their demand, e.g., an embedded controller for small networks to offer low latency and resource utilization. While externally situated controllers can take care of large networks at the price of additional latency. To the best of our knowledge, we are the first to conduct an extensive performance evaluation of controllers for low-power IoT networks. We also share our code~\cite{kulkarni2020thesis} for reproducibility and extension of our findings.



\vspace{-2.00mm}
\section{Background and Related Work} \label{sec:background}

In this section, we provide necessary background on the RPL routing protocol and an overview of SDN implementations for low-power IoT networks.


\boldpar{RPL routing protocol} 
The IPv6-based Routing Protocol for Low-Power-Lossy Networks (RPL)~\cite{RPL-ietf} organizes the routing topology as a tree-like \textit{Destination Oriented Directed Acyclic Graph (DODAG)} that is rooted at the sink or the data collector. Once the root node is established, RPL sends DAG Information Object (DIO) messages towards the other nodes, which carry different routing metrics (e.g., link quality) and an objective function. Each node uses that objective function to select an appropriate parent towards the root and initiate a DAG Advertisement Object (DAO) message to advertise their location to the root. Nodes can also join the network after the topology is formed; in that case, a newly joined node can use DAG Information Solicitation (DIS) messages to solicit a DIO from neighbors. RPL can operate in a \textit{non-storing} mode, where all messages are routed through the root node -- which is the case in our work. 

\boldpar{\usdn} It is a lightweight SDN architecture for embedded \ieee devices. Specifically, the controller sends \usdn Configuration (CONF) messages a RPL DAO from joining nodes -- provisioning that node's flowtable with default rules. Once configured, nodes will periodically send \usdn Network State Update (NSU) messages to the controller to infer their state information (e.g., energy level). On the data plane, \usdn exploits RPL Source Routing Headers (SRH). Specifically, it injects a desired path \textit{directly} into matching packets (as opposed to routing through the root node) and allowing the message utilize regular IPv6 forwarding at intermediate nodes -- reduce the resource consumption and control traffic. Furthermore, \usdn periodically refreshes regularly used rules to avoid repeated queries to the controller. The optimized architecture and protocol stack of \usdn makes it the state-of-the-art software-defined solution for low-power IoT networks. 

\boldpar{\wise} In contrast, the \wise~\cite{gallucio2015sdn} extends the OpenFlow protocol for sensor networks. Unlike traditional stateless data plane devices, \wise sensor nodes maintain three data structures: WISE state arrays, accepted ID arrays, and WISE flow table. The former maintains some state information; the latter two support a chosen set of packet processing and match-actions rules. The protocol stack at each sensor consists of Topology Discovery and Maintenance (TDM),  Packet Forwarding (PF), In-Network Packet Processing (INPP), and Application layers. The first two layers construct and maintain the routing topology and the flow-based packet forwarding, whereas INPP allows a sensor to perform In-Network Packet Processing. The Adaption layer provides an API between a sensor and the controller. However, the stateful \wise design does not offer support for IPv6 and RPL in low-power IoT networks. 


\subsection{Related work} Aside from $\mu$SDN and SDN-WISE, there is now a considerable body of literature examining SDN in IoT. 

SD-WISE ~\cite{anadiotis2015towards} facilitates the heterogeneity characteristic of IoT by leveraging the Network Operating System (NOS), i.e., integrating ONOS controller. SDSense~\cite{haque2019sdsense} delegates some network functions (e.g., neighbor maintenance) to sensors, while the controller takes care of topology discovery and maintenance, etc., that require a global network view. It also provides reliability by adopting the routing topology from \cite{haque2006olear,haque2015selecting}. Another energy-aware software-defined WSN design is presented in~\cite{ding2019interference}. However, none of these solutions is compatible with RPL and IPv6. CORAL-SDN~\cite{theodorou2017coralsdn} proposes an RPL-based topology discovery and maintenance. Recent work in~\cite{saha2020energy} is also built on RPL and Contiki OS to support IPV6 and Network Function Virtualization (NFV), where virtualized data aggregation is optimally deployed in some IoT devices to reduce energy consumption.  
\textit{Crucially, none of these works examine the effect of controller placement (i.e., external vs. embedded) on SDN performance in low-power IoT networks.}

\section{Design and Architecture} \label{sec:approach}

We compare $\mu$SDN and SDN-WISE, two open-source SDN implementations for low-power IoT mesh networks. These implementations each take a different approach to the SDN controller, with $\mu$SDN embedding the controller directly in the mesh while SDN-WISE takes a more traditional route of placing the controller externally on the backend network. In theory, an embedded controller pushed directly onto the mesh should exhibit lower latencies for SDN control messages, while a more powerful controller situated on the backend network should be capable of scaling as the mesh grows. In this regard, to ensure fair comparison between the two architectures, we extend SDN-WISE to support an embedded controller, while we likewise extend $\mu$SDN to support an external ONOS controller.


\begin{figure}[ht]
    \centering
    \vspace{-1.50mm}
    \includegraphics[width=.65\columnwidth]{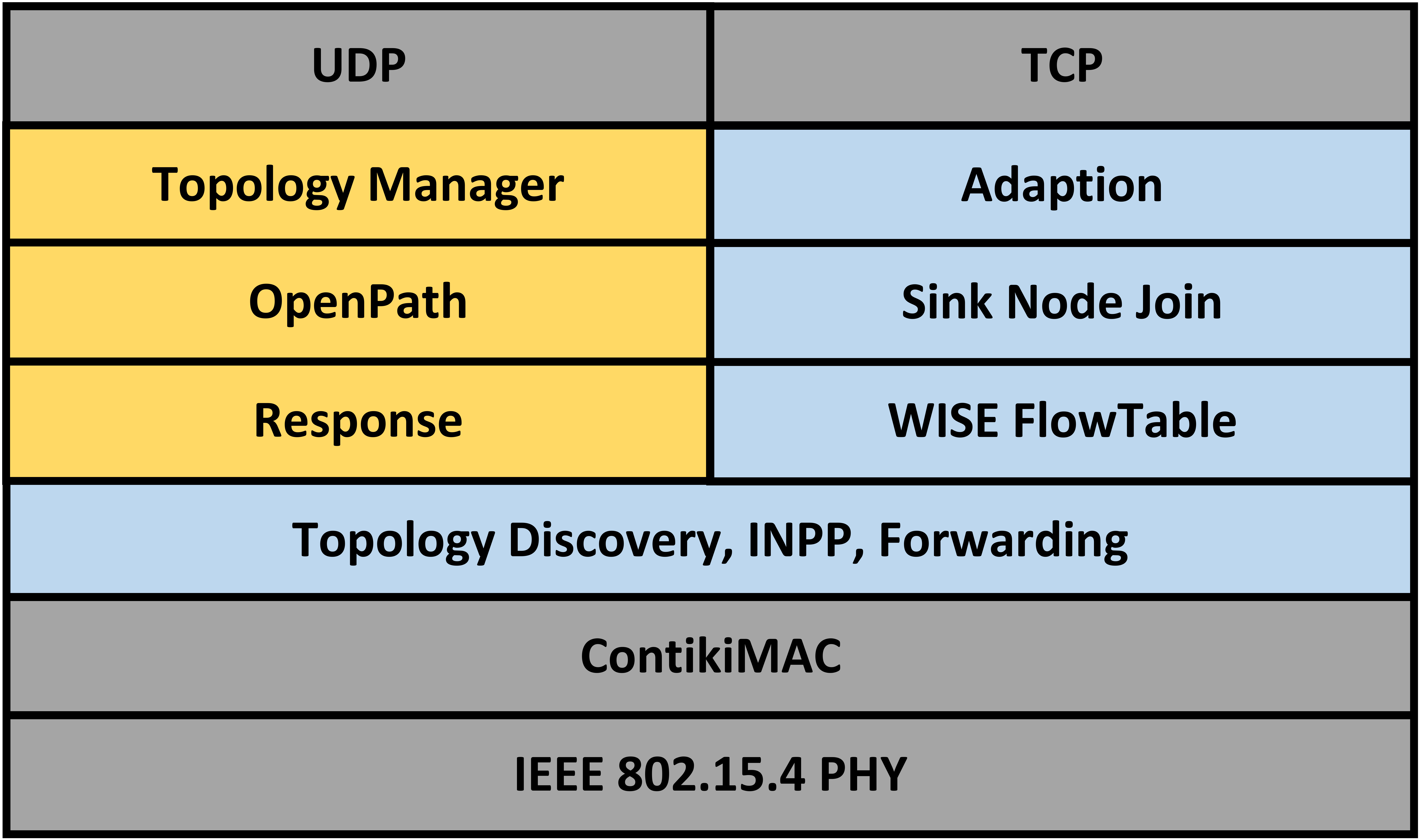}
    \caption{Modified \wise architecture. Yellow components show embedded controller modules, and blue ones are the original \wise node architecture.}
    \vspace{-5.00mm}
    \label{sdnwise-modified}
\end{figure}

\subsection{Integrating Embedded Controller in \wise Architecture}

\usdn introduces its own lightweight embedded controller, \usdn-Atom~\cite{baddeley2020thesis}, which extends the capabilities of a regular low-power node with some limited SDN controller capabilities. \usdn-Atom control nodes are capable of receiving flow requests and responding with flowtable instructions to/from all nodes within its Layer-3 network topology (using RPL), as well as collecting periodic status updates (link quality, neighbors, etc.). This control signaling between the \usdn-Atom controller and \usdn nodes within the IoT mesh network are summarized in Table~\ref{tab:usdn_protocol_signalling}. Since the controller is within the network this allows \usdn-Atom to respond to requests quicker than external controller. In contrast, \wise situates its controller on a backend network outside of the IoT mesh~\cite{gallucio2015sdn} and has recently been extended to support the ONOS SDN Controller~\cite{anadiotis2015towards}. 

\begin{table}[ht]
    \renewcommand{\arraystretch}{.8}
	\centering
	\footnotesize
    \begin{tabular}{l l}
        \toprule
        \textbf{Signaling} & \textbf{Description} \\
        \midrule
        Flowtable Query (FTQ) & Request controller instruction \\
        Flowtable Set (FTS) & Install rule on node \\
        Network State Update (NSU) & Report state to controller \\
        Configuration (CONF) & Initialize joining node \\
        \bottomrule
    \end{tabular}
    \caption{\usdn controller $\leftrightarrow$ node control signaling}
    \vspace{-1.75mm}
    \label{tab:usdn_protocol_signalling}
\end{table}

\begin{table}[ht]
    \renewcommand{\arraystretch}{.8}
	\centering
	\footnotesize
    \begin{tabular}{l l}
        \toprule
        \textbf{Signaling} & \textbf{Description} \\
        \midrule
        Config & Initialize joining node \\
        OpenPath & Collect state information from node \\
        Request & Request controller instruction \\
        Response & Install rule on node \\
        Report & Report state to controller \\
        \bottomrule
    \end{tabular}
    \caption{\wise controller $\leftrightarrow$ node control signaling}
    \vspace{-7.00mm}
    \label{tab:wise_protocol_signalling}
\end{table}

We firstly port \wise's external controller logic to an embedded approach similar to \usdn. Specifically, we  support the controller functions \emph{Topology Manager}, \emph{OpenPath}, and \emph{Response} within a standard \wise low-power node. As in \usdn, these are implemented in the sink node. The \emph{Topology Manager} collects state information (e.g, energy, node position, neighbors) from \wise nodes to construct and maintain the routing topology. The \emph{OpenPath} module helps the controller initiate control packets to those nodes to collect network state information and update the network state view. Finally, the controller uses the \emph{Response} module to send the packets containing flow rules for requesting nodes. Fig.\,\ref{sdnwise-modified} shows the modified architecture for \wise with embedded control logic, while Table~\ref{tab:wise_protocol_signalling} lists the communication messages between the \wise controller and nodes. 


\subsection{Integrating ONOS Controller to \usdn Architecture}

Likewise, we extend the \usdn stack to support an external SDN controller ONOS -- creating ONOS components to support \usdn control functions externally rather than on the embedded \usdn-Atom controller. 

\begin{figure}[ht]
    \centering
    \includegraphics[width=.75\columnwidth]{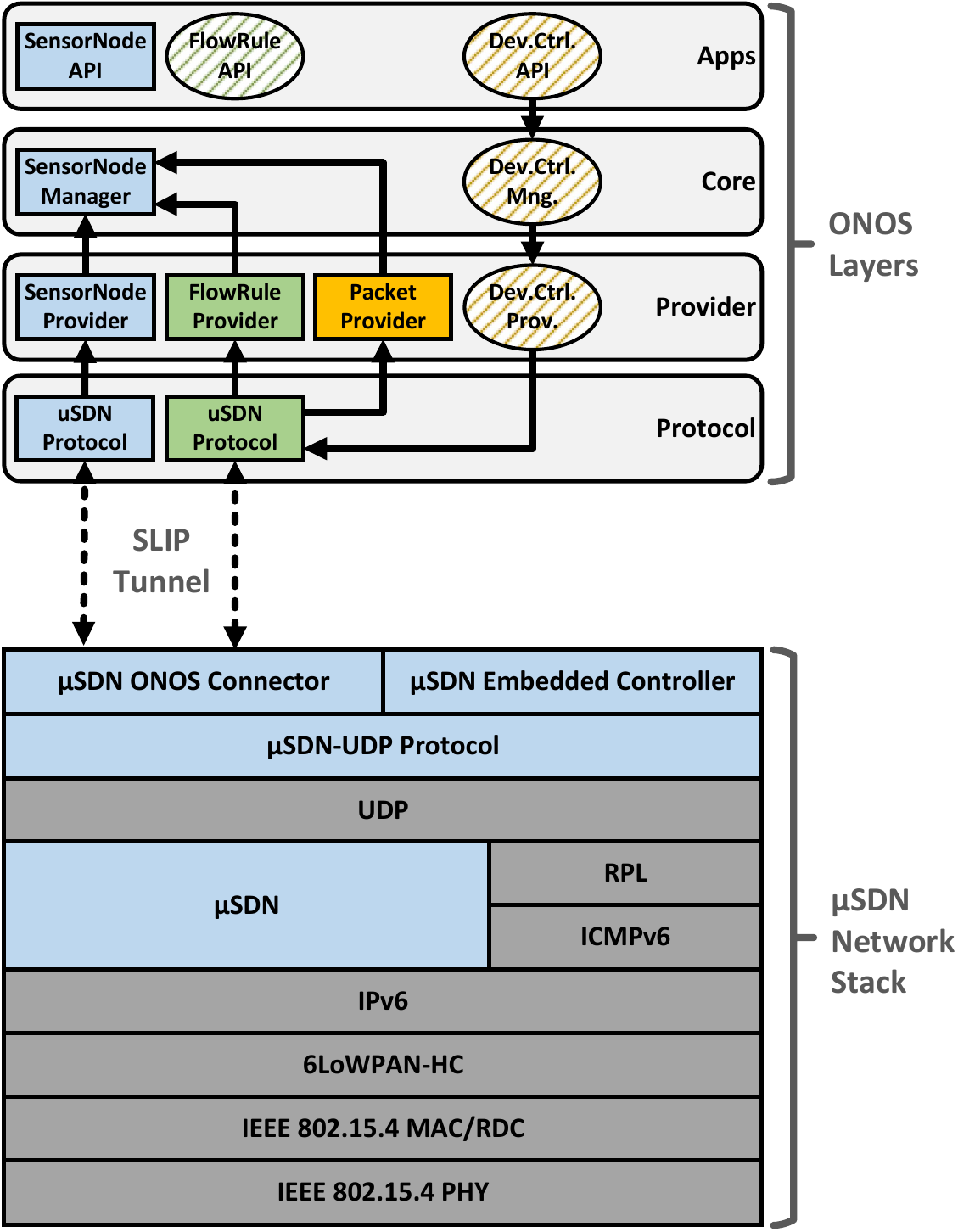}
    \caption{Extension of \usdn to the ONOS controller architecture.}
    \vspace{-3.00mm}
    \label{fig:usdn_onos_extension}
\end{figure}

\boldpar{Adaption of ONOS Subsystems} 
A number of different ONOS components were developed. Fig.\,\ref{fig:usdn_onos_extension} shows rectangular components designed as per the \usdn-Atom controller functionality and hatched ones extended to support the \usdn message format. Each subsystem is marked in the same color, while the arrows demarcate the directions in which messages travel within the ONOS controller. 

\emph{SensorNode Subsystem} -- when ONOS receives a message at the protocol layer, an event is triggered that notifies the SensorNode provider that a node has joined. This information is then propagated to the SensorNode manager that maintains node information (e.g., NodeID). The SensorNode API offers access to this information. \emph{FlowRule Subsystem} -- FTQ messages from \usdn nodes are sent to the FlowRule Provider. This wraps the FTQ in the ONOS format. The corresponding FTS message is reformatted back to the \usdn protocol and sent back to the mesh. The FlowRule API is designed to support this format conversion. \emph{Packet Subsystem} -- ONOS needs to maintain topology and relies on \usdn NSU messages to achieve this. The Packet subsystem is designed for the Provider layer that receives NSUs and generates the corresponding CONF messages. The Packet provider also offers the format conversion between ONOS and \usdn before passing them to an appropriate subsystem. \emph{DeviceControl Subsystem} -- depending on the required device configuration for the underlying low-power sensors, this subsystem generates appropriate messages like the device on/off. These messages are then converted to the right format while they travel between the ONOS and \usdn layers. 

\boldpar{Workflow of the \usdn-ONOS Architecture} 
\usdn traffic is now handled by ONOS in three stages. \emph{Controller Discovery and Join} -- firstly, the \usdn sink receives DAOs from nodes that have joined the RPL DAG and forwards these to ONOS. These flow to the SensorNode subsystem that stores and maintains the topology. The Packet subsystem then generates a CONF message to set the nodes with flowtable rules, timers, etc. After converting to \usdn format, these are sent back to the sink to forward over RPL. \emph{Node Updates} -- once configuration is complete, nodes are ready to communicate, i.e., convey data plane traffic to the sink. However, the ONOS controller needs to know the current network state. Accordingly, \usdn nodes periodically send NSUs with configured metrics (energy, neighbors, etc.) to the controller. The Packet subsystem receives these NSUs and updates the topology in the SensorNode subsystem. The DeviceControl subsystem actively keeps track of these metrics and sets priorities on the information that the controller must aware of.
\emph{Controller Response} -- in addition to the periodic control message sharing, nodes also need to generate on-demand control traffic. For example, \usdn nodes send an FTQ message to the controller if it misses a flow-rule entry. The FTQ message is handled by the FlowRule subsystem, where the FlowRule provider generates an FTS message that contains appropriate flow rules for the requesting node. 

\section{Evaluation Setup} \label{sec:setup} 

In this section, we outline the evaluation setup. Our experiments run on a machine with a 3.8 GHz 6 core CPU and 16 GB RAM equipped with Cooja 2.7 simulator for Contiki OS. \usdn nodes consist of a MSP430F5438 platform with a CC2420 radio. \wise nodes consist of EMB-Z2530PA based sensor nodes. Table \ref{tab:ProtocolStack} describes protocol stacks of \usdn and \wise used in the evaluation. In both cases, the nodes form 50 node grid or random topologies, where the sink node has the embedded customized SDN controller for low-power IoT networks. 

\begin{table}[ht]
    \renewcommand{\arraystretch}{1.0}
	\centering
	\footnotesize
    \begin{tabular}{L{1cm} c c}
        \toprule
        \textbf{Layer} & \textbf{\usdn} & \textbf{\wise} \\
        \midrule
        Control & Embedded & Embedded \\
        Network & IPv6 + RPL & TD INPP Forwarding \\
        MAC & CSMA/CA  & CSMA/CA \\
        RDC & Contiki MAC & Contiki MAC \\
        PHY & IEEE 802.15.4 & IEEE 802.15.4 \\
        \bottomrule
    \end{tabular}
    \caption{\usdn and \wise protocol stacks}
    \label{tab:ProtocolStack}
\end{table}


We also connect an ONOS controller to the \usdn architecture to compare the performance of the external and embedded controllers. In this setup, we use a SLIP connection between the sink and ONOS. The sink node acts as a border router to facilitate the connection between \usdn nodes and the ONOS controller.  We repeat each evaluation 50 times to take the average with a 95\% confidence interval. Table \ref{tab:Cooja_Parameter} presents various parameters and their settings in the Cooja simulator that we use in the evaluation.

\begin{table}[ht]
    \renewcommand{\arraystretch}{1.0}
	\centering
	\footnotesize
    \begin{tabular}{L{3cm} c}
        \toprule
        \textbf{Parameter} & \textbf{Value}  \\
        \midrule
        Duration & 5 min \\
        Transmission Range & 50m\\
        Interference Range & 50m\\
        Max Bit rate & 9 bit/sec \\
        Max Flow request rate & 102 flows/sec \\
        Link Quality & 90\% \\
        Radio Medium & UDGM (distance loss) \\
        RPL mode & Non-storing mode\\
        RPL Route Lifetime & 5min\\
        Flowtable Lifetime & 5min\\
        \bottomrule
    \end{tabular}
    \caption{List of Cooja parameters used in the evaluation.}
    \vspace{-5.00mm}
    \label{tab:Cooja_Parameter}
\end{table}

\section{Results Discussion} \label{sec:results}

This section first presents a detailed discussion on the \usdn and \wise architecture comparison results. Then, we discuss embedded and external controller comparison.

\subsection{Node Architecture Comparison}
As part of this comparison, we study the communication energy consumption, delay, and packet delivery ratio (PDR) over varying load and route length to infer the superiority of one architecture over the other. 

\begin{figure}[t]
\begin{subfigure}[t]{.49\columnwidth}
    \includegraphics[width=\columnwidth]{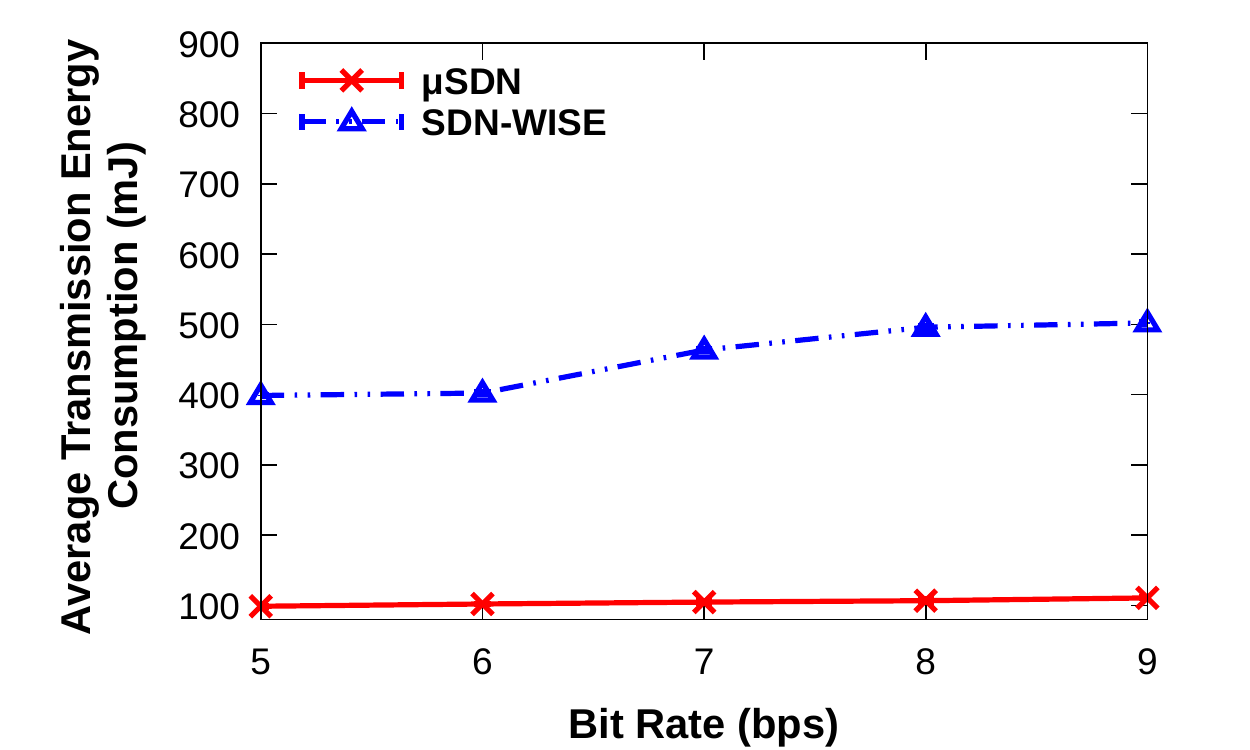}
    \subcaption{Bit Rate}
    \label{fig:energy_grid_bit}
\end{subfigure}
\begin{subfigure}[t]{.49\columnwidth}
    \includegraphics[width=\columnwidth]{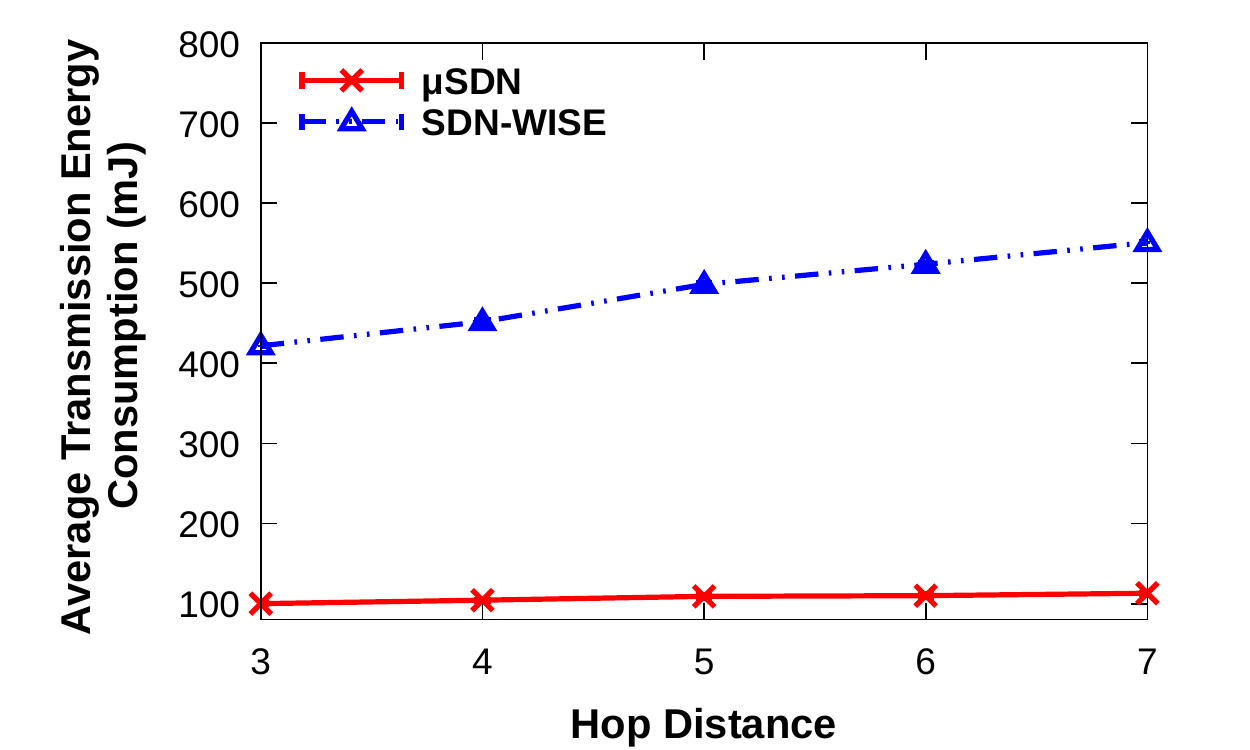}
    \subcaption{Hop Distance}
    \label{fig:energy_grid_hop}
\end{subfigure}
\caption{Mean energy consumption in a grid topology.}
\vspace{-5.00mm}
\label{fig:energy_grid}
\end{figure}

\boldpar{Communication energy consumption} The average energy consumption from Fig.\,\ref{fig:energy_grid} depicts that \usdn has a significantly lower consumption compared to \wise. \usdn is built on the standard \ieee low-power IoT protocol stack and (as previously describes) contains optimizations for SDN control traffic. Also, \usdn is a stateless architecture, where relay nodes do not maintain any routing tables. Finally, it deploys interference-aware routing that reduces the number of re-transmissions. All these features contribute to the significant energy savings of \usdn, which is not the case for the stateful \wise. We observe a similar performance trend with varying load and path length, where route length has more impact on \wise. The behavior that again can be explained as the lack of interference-aware packet forwarding. 

\begin{figure}[t]
\begin{minipage}[t]{0.49\columnwidth}
    \includegraphics[width=\linewidth]{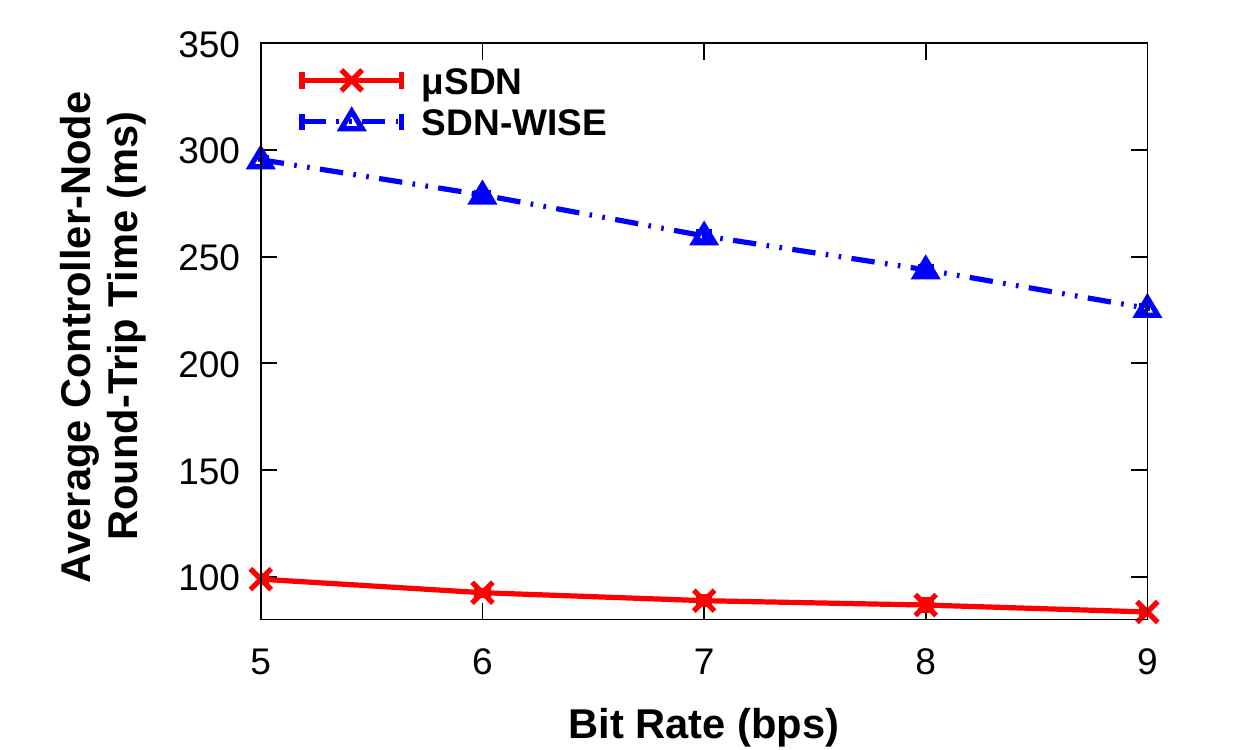}
\end{minipage}
\begin{minipage}[t]{0.49\columnwidth}
    \includegraphics[width=\linewidth]{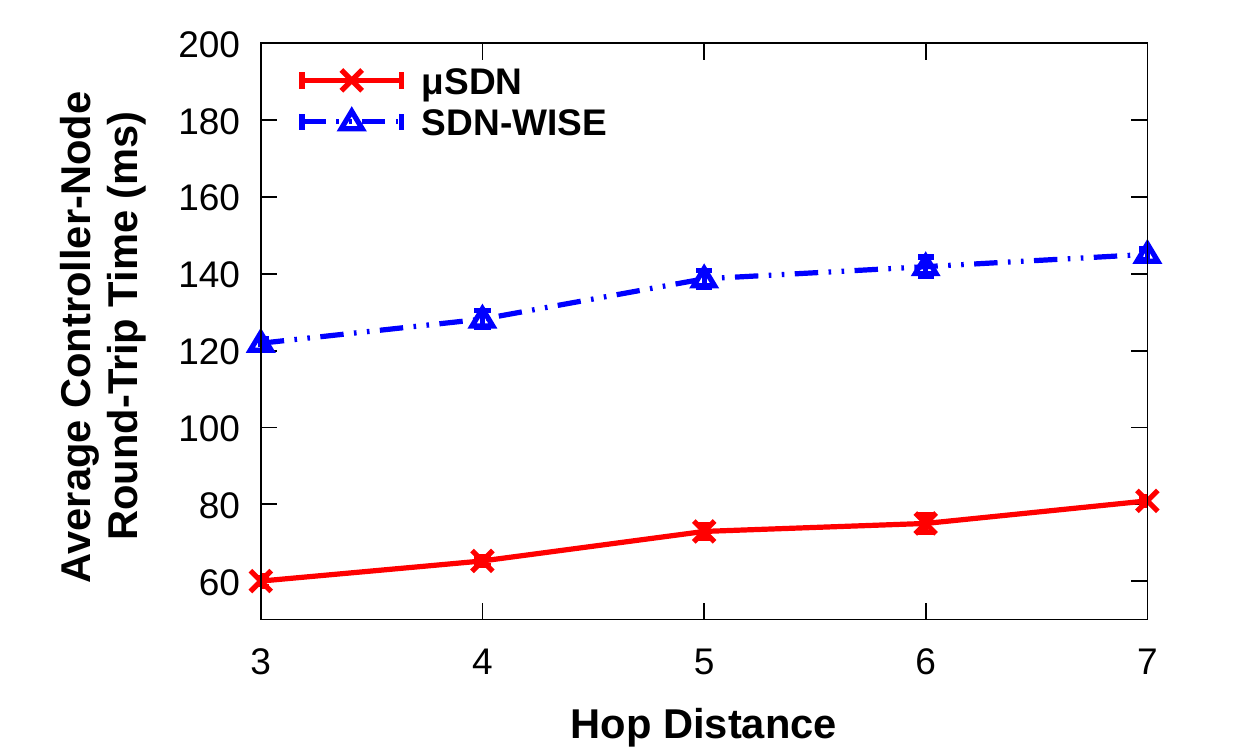}
\end{minipage}
\caption{Mean controller-node delay in a grid topology.}
\vspace{-5.00mm}
\label{rtt_grid}
\end{figure}

\boldpar{Flow-rule installation time} As seen in Fig.\,\ref{rtt_grid}, \usdn shows better performance when we measure the node-to-controller round-trip time of flow rule installation requests. \usdn has better performance due to using source routing at the relays. Their number is usually much higher compared to sources. However, relays in \wise also maintain flow tables that impact the rule installation time. As the bit rate increases, we see a decrease in that time as packets reach destinations quickly. Nonetheless, the time increases with the increasing hop-distance due to the chance of interference and additional requests to the controller. 


\begin{figure}[ht]
\begin{minipage}[t]{0.49\columnwidth}
    \includegraphics[width=\linewidth]{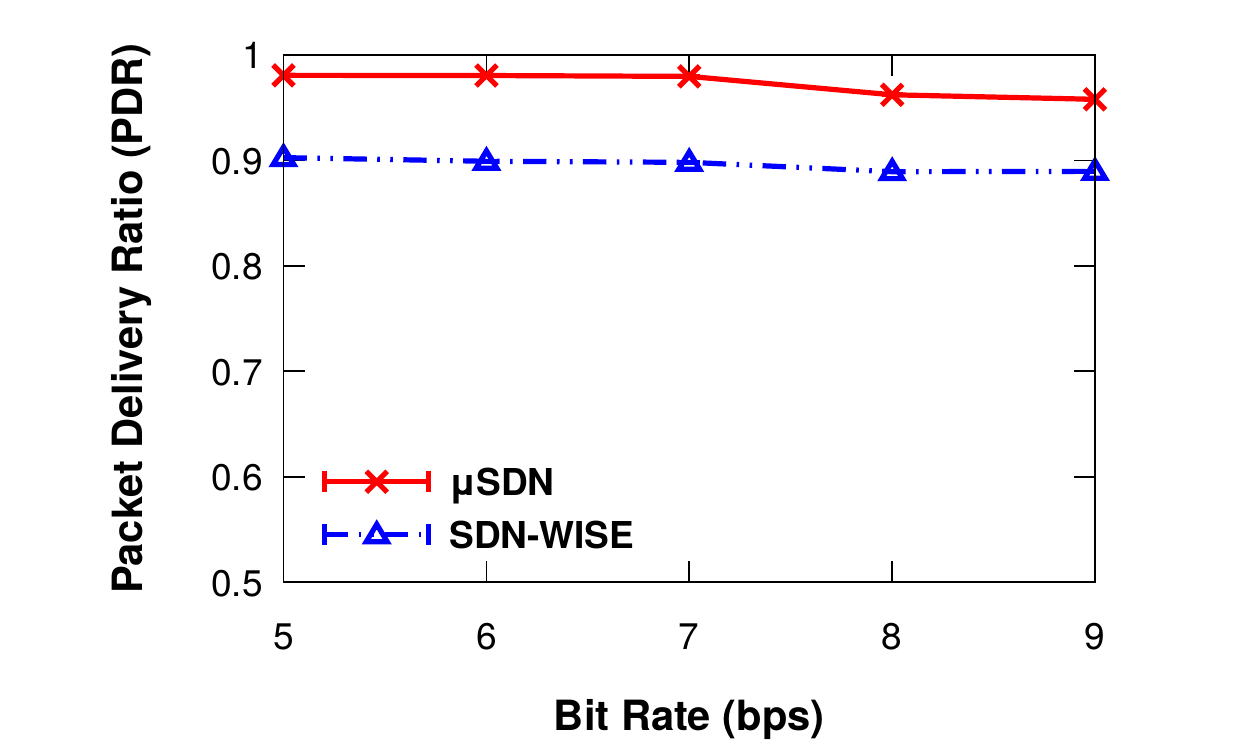}
\end{minipage}
\begin{minipage}[t]{0.49\columnwidth}
    \includegraphics[width=\linewidth]{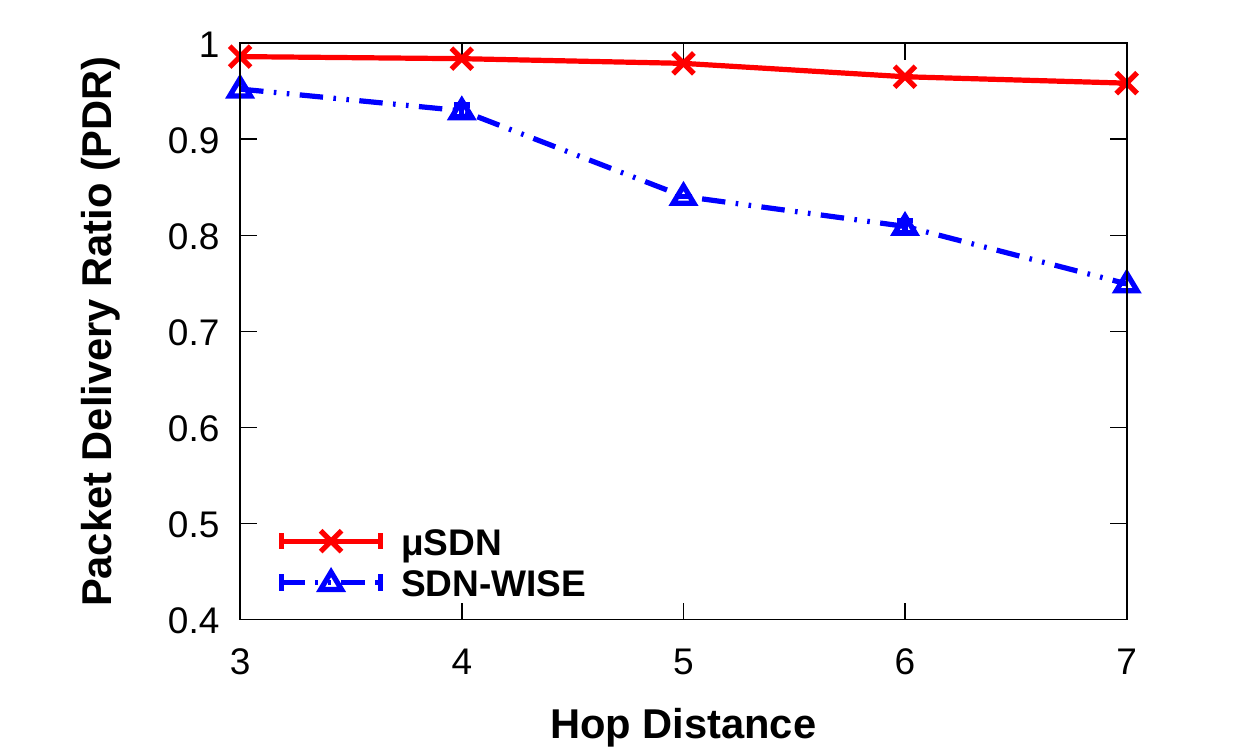}
\end{minipage}
\caption{Mean PDR in a grid topology.}
\label{pdr_grid}
\end{figure}

\boldpar{Packet reliability} The average Packet Delivery Ratio (PDR) for both architectures is presented in Fig.\ref{pdr_grid}. \usdn has a higher PDR than \wise, as expected. The optimization across the protocol stack, interference-aware source routing all contribute to this performance of \usdn. For example, it forwards packets over alternative routes in the presence of interference instead of packet drop or retransmission in \wise. Thus, building the \usdn architecture on top of a standard protocol stack with the required optimization for software-defined networks brings considerable advantages. In the following, we measure its performance with an embedded (part of the IoT networks) and standard external controller (ONOS) to find the appropriate controller choice for the low-power IoT networks. 



\subsection{Controller Comparison}
This section compares the performance of embedded (\usdn) and external (ONOS) controllers considering standard evaluation metrics like throughput, latency, topology discovery, and update time. In the rest of this section, we present the results for the grid topology as the performance trend is similar on the random one.

\begin{figure}[ht]
\begin{minipage}[t]{0.49\columnwidth}
    \includegraphics[width=\linewidth]{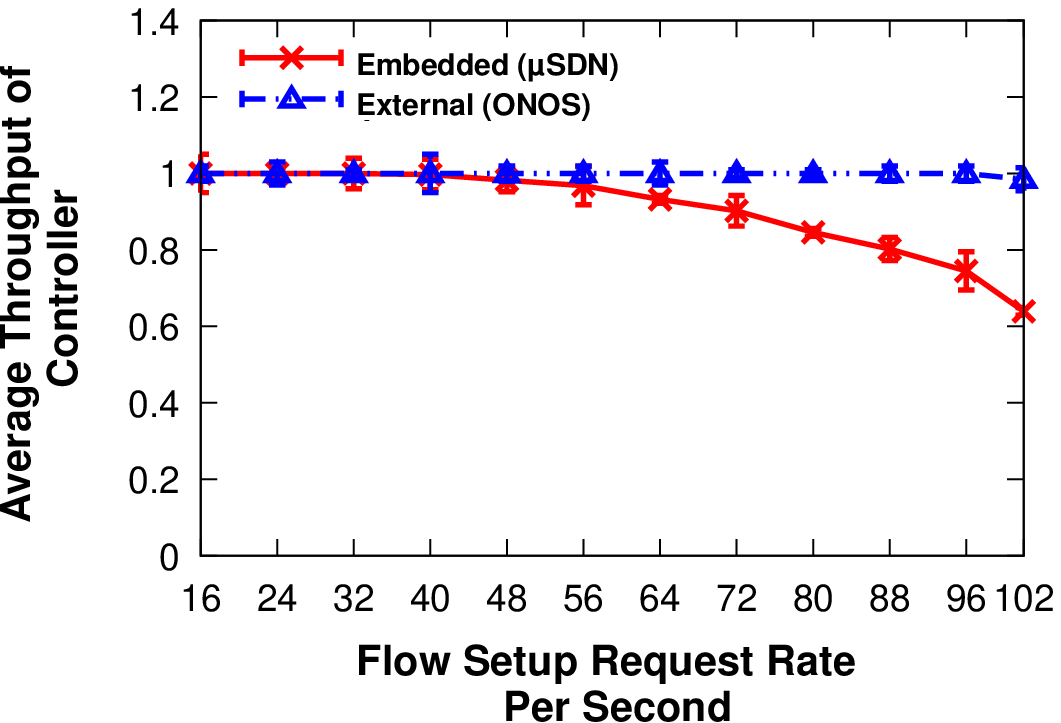}
\end{minipage}
\begin{minipage}[t]{0.49\columnwidth}
    \includegraphics[width=\linewidth]{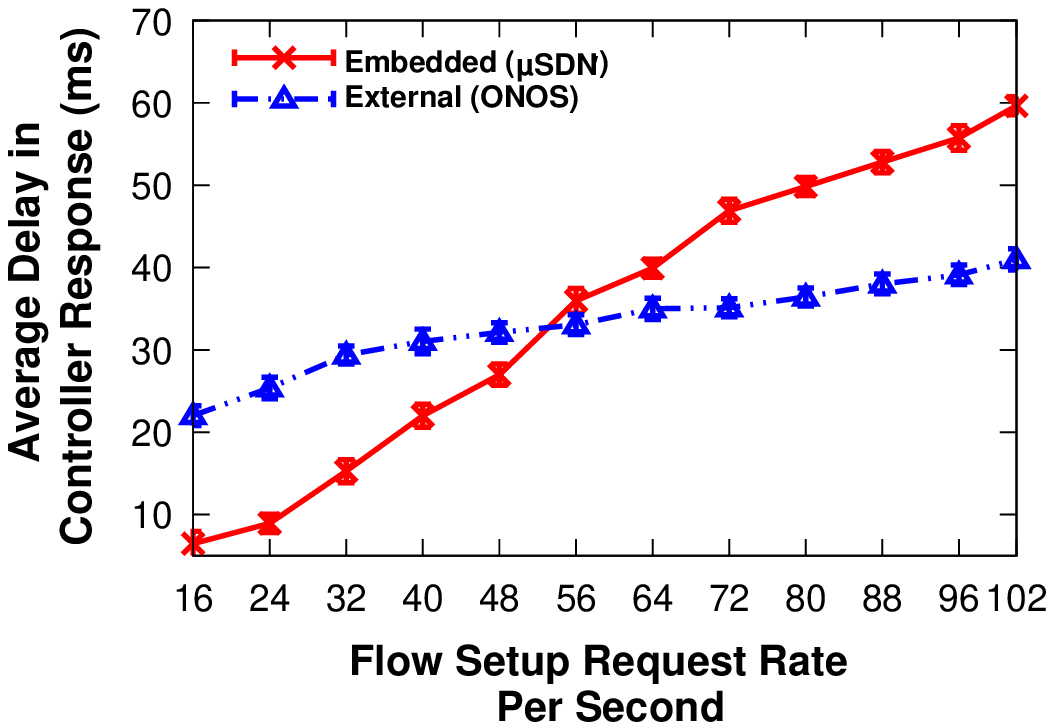}
\end{minipage}
\caption{Mean throughput and latency.}
\vspace{-5.00mm}
\label{throughput_latency_grid}
\end{figure}

\begin{figure}[ht]
\begin{minipage}[t]{0.49\columnwidth}
    \includegraphics[width=\linewidth]{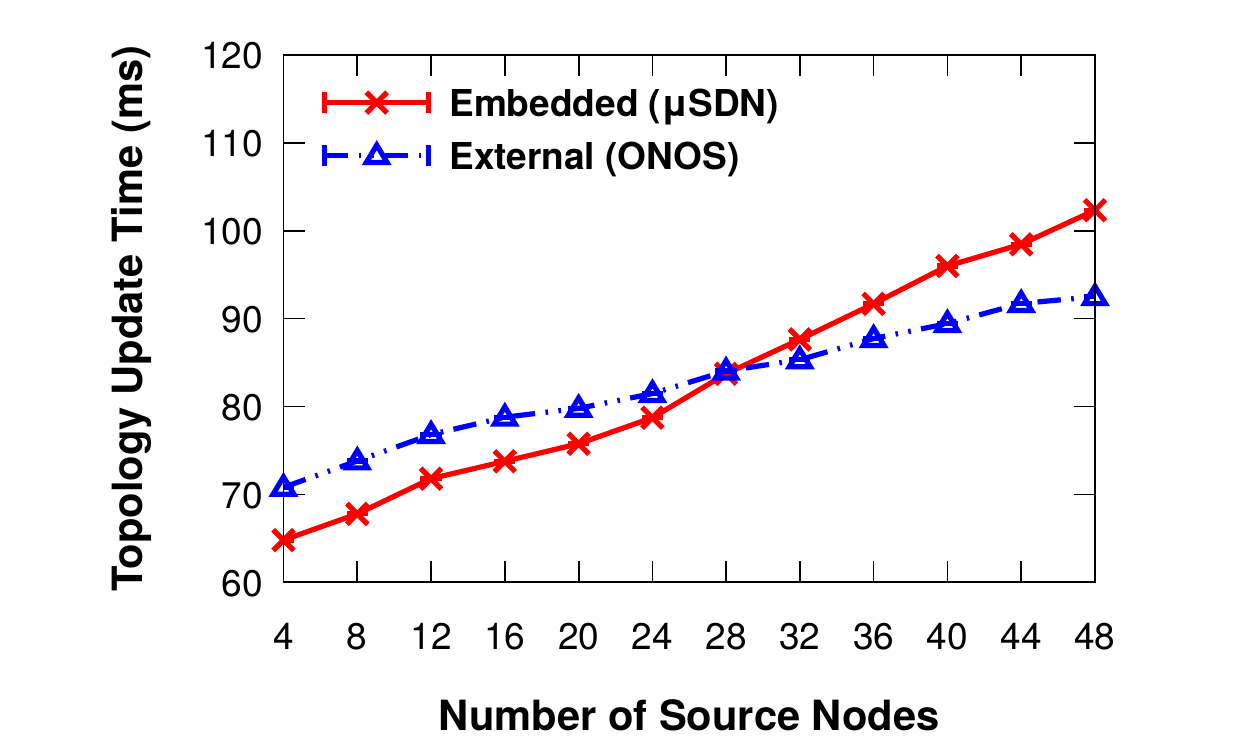}
\end{minipage}
\begin{minipage}[t]{0.49\columnwidth}
   \includegraphics[width=\linewidth]{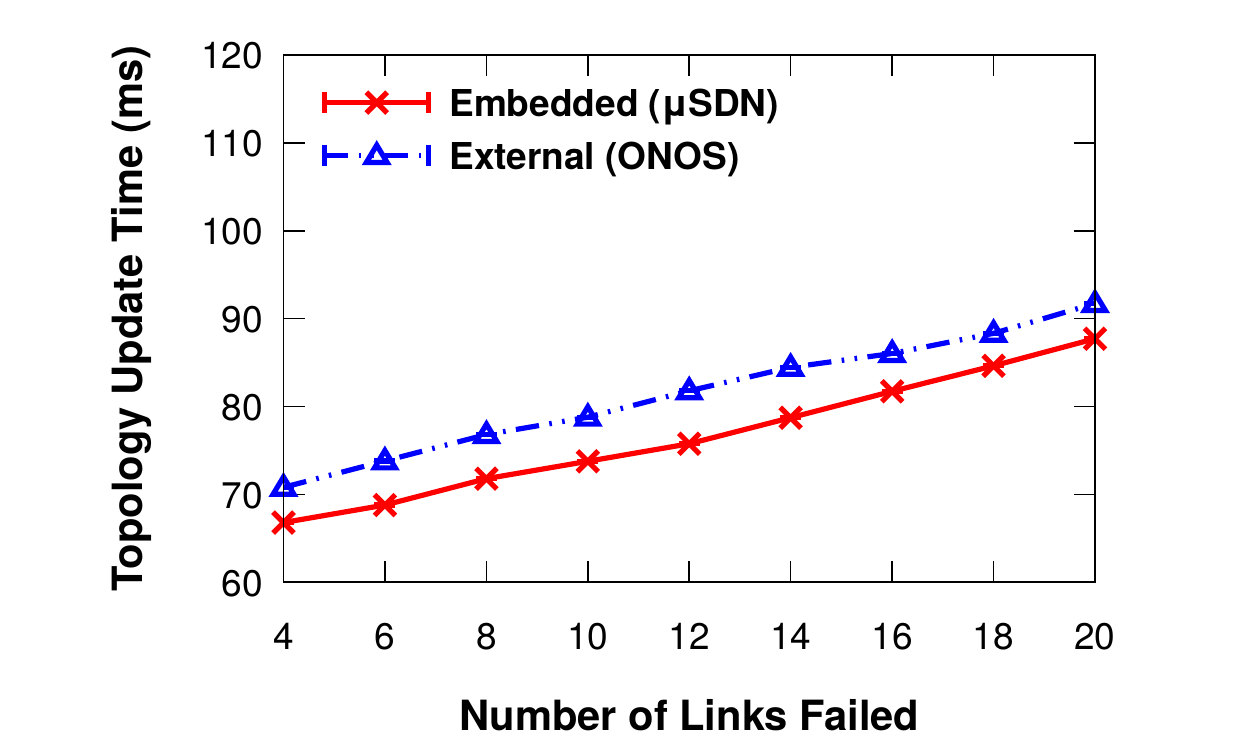}
\end{minipage}
\caption{Mean topology update time for source and link failures.}
\vspace{-5.00mm}
\label{topo_update_source}
\end{figure}

\boldpar{Throughput and latency measurement} Our goal in this evaluation is to check the performance of the chosen controllers in terms of handling a varying number of flow-rule installation requests per unit time from all sensor nodes, i.e., we consider a highly loaded network. The throughput is presented in Fig.\,\ref{throughput_latency_grid}(a). We observe an interesting performance trend. Both controllers have the same throughput for up to 50 requests per unit time. The performance of the embedded  \usdn Atom controller starts degrading after that point with the increasing number of requests. The standard distributed ONOS controller's processing capability takes the lead under high load, which is not the case for the domain-specific embedded one. We observe a similar performance trend in the case of latency measurement. \usdn`s embedded controller latency in Fig.\,\ref{throughput_latency_grid}(b) is significantly better than the external ONOS controller for a request rate up to 50. After that point, it shows the opposite trend compared to ONOS. The embedded controller can process all requests within its resource budgets and get back to the nodes quickly at a low rate. However, ONOS suffers from the communication delay as it is external to the network. As the rate increases, communication latency is compensated by the processing capability of ONOS, where the embedded controller falls back.

\boldpar{Topology discovery and update time} 
We measure the topology discovery time for small to large topologies, where ONOS takes a constant amount of additional time compared to the embedded controller. This result is expected as the embedded one is part of the IoT network and works on the same protocol stack that resides in the IoT nodes. However, the gateway or sink node connects to the external ONOS controller that furthermore requires protocol compatibility. Note that due to the space limitation we are not presenting the discovery time comparison. Topology updates are required because of a change in the network due to the failure or unavailability of nodes or links. In this experiment, we randomly fail a fixed number of links from the network for varying sources and measure the topology update time in two settings: with and without data traffic flowing in the network. In both cases, we observe a performance trend (Fig.\,\ref{topo_update_source}(a)) similar to that of the latency, i.e., the embedded controller takes less time while the number of sources is small. As this number increases, ONOS takes the lead over the embedded one. One the other hand, when we vary the number of failed links for a given number of sources in Fig.\,\ref{topo_update_source}(b), we observe that the update time increases with the increasing number of failed links. Thus, the number of sources or the network size has the most significant impact on the performance of the controllers. 

\vspace{-1.00mm}
\section{Conclusions} \label{sec:conclusion}



This paper has performed an extensive evaluation of two SOTA low-power IoT network architectures: \wise and \usdn. We have found that the control-layer optimizations present in the \ieee and IPv6 compliant \usdn allow it to outperform \wise across energy, latency, and reliability metrics. However, we have shown that while the embedded controller approach in \usdn initially demonstrates significant performance advantages, as the network scales this advantage is reduced. Crucially, we can quantify a `crossover point' of around 50 nodes / requests-per-second at which an external controller with greater resources (such as ONOS) is better positioned to handle larger or more demanding networks: thus a controller should be chosen based on applications' QoS.

\vspace{-1.00mm}



\bibliographystyle{IEEEtran}
\bibliography{references}

\end{document}